\documentclass{article}

\usepackage{arxiv}
\usepackage{algorithm}
\usepackage{algorithmic}
\usepackage{times}  
\usepackage{amsmath,amsfonts,amssymb,amsthm}
\usepackage[utf8]{inputenc} 
\usepackage[T1]{fontenc}    
\usepackage{hyperref}       
\usepackage{url}            
\usepackage{booktabs}       
\usepackage{amsfonts}       
\usepackage{nicefrac}       
\usepackage{microtype}      
\usepackage{lipsum}		
\usepackage{graphicx}

\theoremstyle{definition}
\newtheorem{definition}{Definition}
\title{Boosting Retailer Revenue by Generated Optimized Combined Multiple Digital Marketing Campaigns}

\author{ {\hspace{1mm}Yafei Xu, Tian Xie, Yu Zhang} \\
	Alibaba Group\\
	Jiancaichen Middle Road No.27\\
	100027, Beijing, PR China \\
	\texttt{\{xyf252918,\ xietian.xt,\ zy223687\}@alibaba-inc.com} \\
}

\renewcommand{\shorttitle}{}

\hypersetup{
pdftitle={Boosting Retailer Revenue by Generated Optimized Combined Multiple Digital Marketing Campaigns},
pdfauthor={Yafei Xu},
pdfkeywords={First keyword, Second keyword, More},
}

\begin{document}
\maketitle

\begin{abstract}
Campaign is a frequently employed instrument in lifting up the GMV (Gross Merchandise Volume) of retailer in traditional marketing. As its  counterpart in online context, digital-marketing-campaign (DMC) has being trending in recent years with the rapid development of the e-commerce. However, how to empower massive sellers on the online retailing platform the capacity of applying combined multiple digital marketing campaigns to boost their  shops' revenue, is still a novel topic. In this work, a comprehensive solution of generating optimized combined multiple DMCs is presented. Firstly, a potential personalized DMC  pool is generated for every retailer by a newly proposed neural network model, i.e. the DMCNet (Digital-Marketing-Campaign Net). Secondly, based on the sub-modular optimization theory and the DMC pool by DMCNet, the generated combined multiple DMCs are ranked with respect to their revenue generation strength then the top three ranked campaigns are returned to the sellers' back-end management system, so that retailers can set combined multiple DMCs for their online shops just in one-shot. Real online A/B-test shows that with the integrated solution,  sellers of the online retailing platform increase their shops'  GMVs with approximately 6$\%$.
\end{abstract}
\section{1. Background}
\begin{figure*}[t]
\centering

\includegraphics[width=0.6\textwidth]{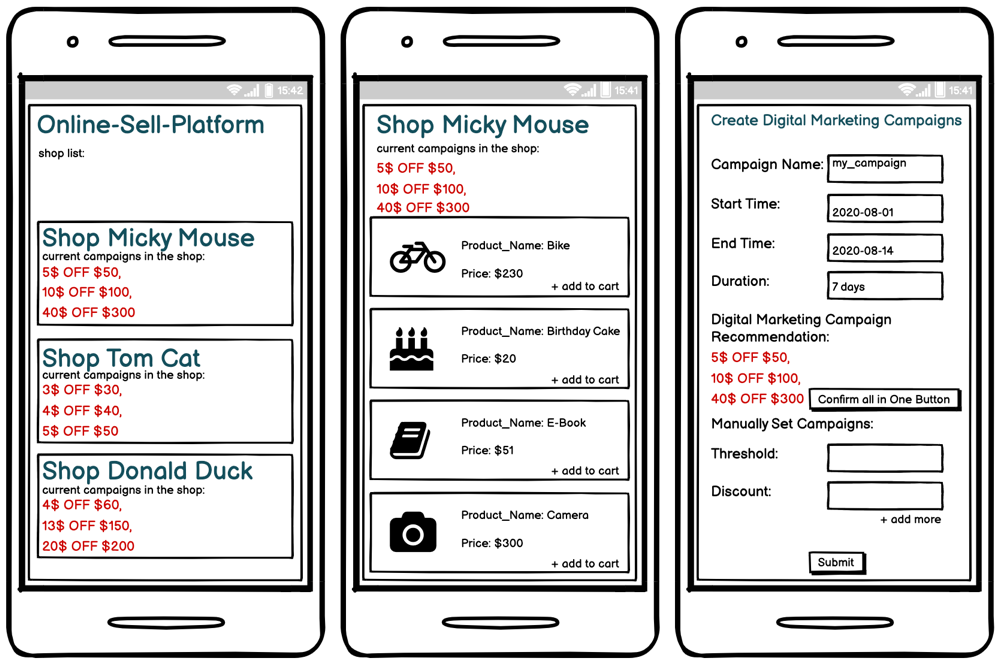}
\caption{The leftmost plot illustrates the online-sell-platform with three independent online-shops, in which there list distinct current DMCs in red in the shop. The middle plot gives the consumer side interface with red fonts indicating the current DMCs active in the shop. The rightmost plot shows the interface of the retailer side, setting DMCs with the red fonts giving the multiple DMCs recommendation.  }
\label{iphone}
\end{figure*}

In China, with the exploding evolution of e-commerce, online-sell-platform has been iterated  qualitatively and quantitatively to a highly competitive period. Therefore, in order to survive in such a dynamic and competition-fierce market environment, every online-sell-platform  has to innovate new techniques to attract more consumers in buyer side and increase more retailers' revenue in seller side. This work solves a concrete problem, i.e. how  to help retailers (shops) to lift up revenue by employing DMCs. As illustrated in Figure \ref{iphone}, each online shop (on the leftmost plot) on the online-sell-platform has given their DMCs in red. And  in the rightmost plot each shop can complete the DMC settings manually or automated by using DMC recommendation indicated in red. The DMC recommendation is all what this paper concerning about. Though given the prototype plots in Figure \ref{iphone}, 
as a leading online-sell-platform in China the real application has been implemented  by authors before this paper.

\section{2. Introduction}
In traditional marketing, campaign is often used by retailers to lift up revenues. In recent years, with the rushing development of e-commerce, DMC has been in tendency in online marketing. Especially, with the flexible implementation of combination of DMCs, more revenue can be achieved compared with those not-combined DMCs. For example, campaign threshold-discount pairs $<threshold,discount>$, say $<60, 5>$ and $<70, 8>$, mean that in an online shop a consumer can have two campaigns at the same time. If the consumer buys a basket of food worth $\$60$ then the campaign $<60, 5>$ is triggered, and the consumer obtains $\$5$'s discount. If the consumer buys a basket of food worth $\$70$ then the both campaigns $<60, 5>$ $<70, 8>$ are triggered, and the consumer obtains $\$8$'s discount. Applying these DMCs is of benefit to transfer the low-value consumer to high-value consumer, which will finally result in boosting revenue. In traditional retailing times when the big data infrastructure is not accessible, sellers use their selling experience frequently. However in new retailing era, with the help of big data, online-sell-platform can help retailers to increase their return by automated generating combined multiple DMCs. This paper presented a comprehensive solution to generating combined multiple DMCs.

Here it is necessary to introduce the  rules of DMC. It can be understood as a special kind of digital coupon. That is the DMC is issued  by individual online shops, hence each shop on the online-sell-platform has its own DMCs oriented to distinct marketing objectives. And each DMC has a trigger threshold and a discount amount, i.e. DMC threshold-discount pair. For example, an online shop issues a DMC with a threshold-discount pair of $<90, 10>$, then if a consumer's shopping cart is fulfilled with products worth more than $\$90$ then the $<90, 10>$ DMC is triggered. And the consumer will obtain a $\$10$ discount, i.e. the consumer saves $\$10$ and only needs to pay $\$80$.

In fact, it is a rather complicated problem to assure the effectiveness of recommending the digital marketing campaigns. From the macro aspect, the price elasticity can be depicted as the relationship between the discount strength and the selling volume, it is a multi-factor problem and it is hard to collect data to compute price elasticity. Therefore the problem is difficult to solve with price elasticity method commonly used in marketing field. 

    This paper proposed a comprehensive solution to solve such problem based on DMCNet and Randomlized USM (unconstrained submodular maximization) Algorithm\  {\cite{buchbinder2015sb1}}. DMCNet is a neural network model proposed to calculate the DMC's triggering probability for each consumers. More importantly, the revenue of each DMC for a online shop can be calculated by DMCNet. More detail will be shown in Section 3. 
    
    Randomlized USM Algorithm proposed a solution to non-monotonic submodular problem like optimal combination of DMCs. Together with DMCNet, the maximum of DMCs revenue will be obtained.

This paper contains four parts, In section $3$, DMCNet  will be introduced. Section $4$ describes the motivation to resolve optimal combination of DMCs as a non monotonic submodular problem, and the method to obtain optimal revenue of this problem. The whole recommendation solution and pseudo code are presented in section $5$. Experiment will be given in section 6. Last section concludes the paper.



\begin{figure*}[t]
\centering
\includegraphics[width=.7\textwidth]{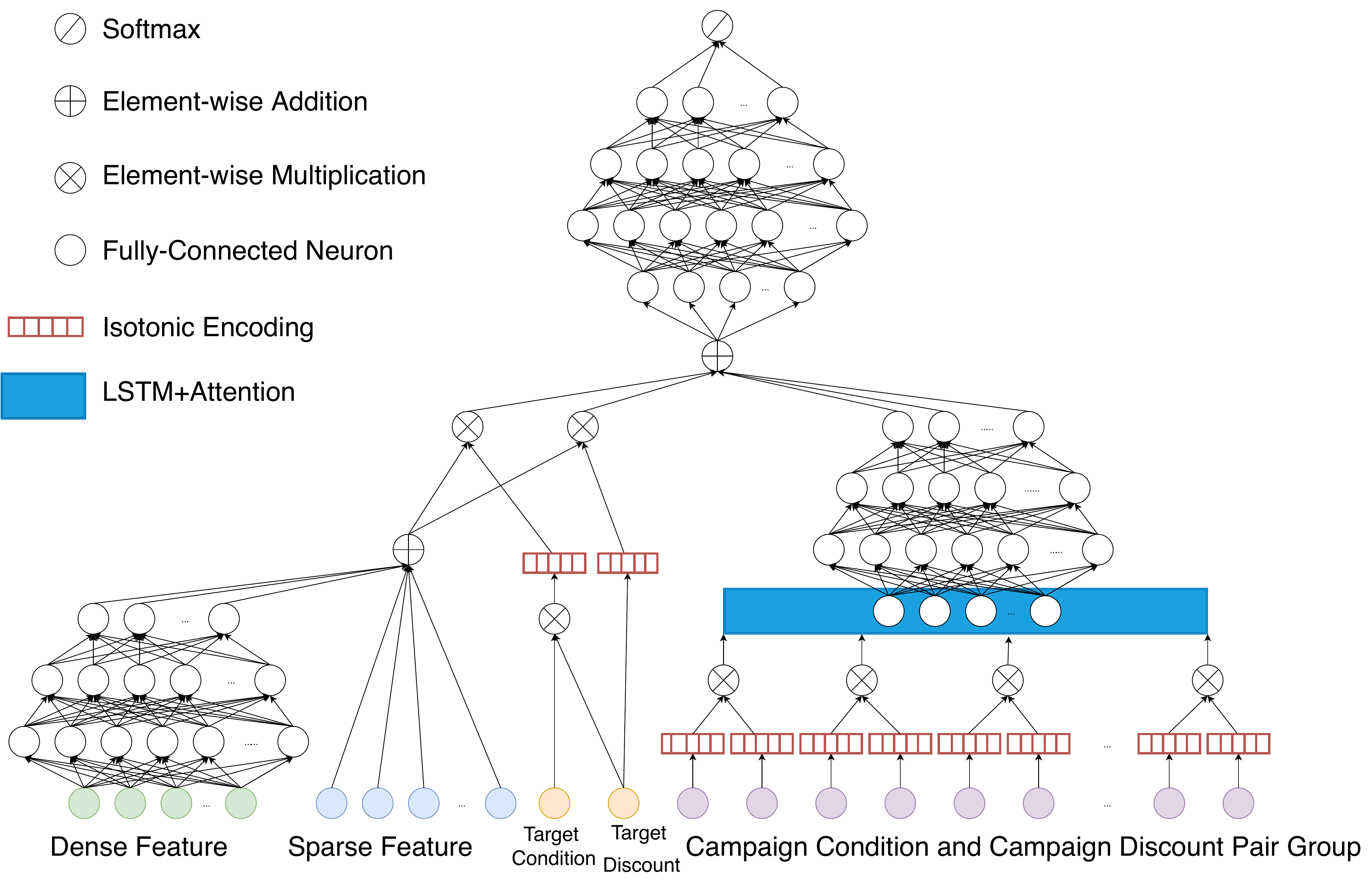} 
\caption{Topological structure of  DMCNet}.
\label{figDMCNet}
\end{figure*}




\subsection{Review of Related Techniques}
The main techniques invested in this work contain recommender system (RS), deep learning (DL).

Recommender system is a  tool to retrieve information to help users make decisions efficiently. Especially with the fast development of e-commerce, RS plays a pivotal and indispensable role, not only in consecutively ameliorate the consuming experience but also helpful in lifting up retailers' revenue. Essentially, a typical RS leverages user profile, item features, user-item interaction information and other inputs like temporal and spatial data to predict the user's next choice. Typical RS can be categorized as three classes, i.e. collaborative filtering,  content-based RS and hybrid RS, see \cite{rscategory}.

Deep learning has been receiving in decades more and more attentions after tremendous advance in theory \cite{lecun1989cnn0,lecun1998cnn1,hinton2009DBN,cho2014gru} and computing  infrastructure \cite{dean2008mapreduce,shvachko2010hadoop,zaharia2011spark,jouppi2017TPU}. Massive applications have been innovating in industry and achieving in SOTA, e.g. in fields of natural language processing \cite{vaswani2017attention, devlin2018bert,yang2019xlnet} and computer vision \cite{simonyan2014vgg,he2016resnet,redmon2016you}. The hybrid model incorporating deep learning into recommender system has become the tendency  in industrial recommendation system \cite{zhang2019rssurvey}, e.g. deep collaborative model \cite{li2015deepcf}, wide and deep model \cite{cheng2016wide}, deepfm model \cite{guodeepfm}, deep interest model \cite{zhou2018din,zhou2019dien}. 


Here it is necessary to mention that the comprehensive solution in this work lends the idea from deep-learning based recommender system and combinatorial optimization. 

\section{3. Digital-Marketing-Campaign Net}
In this paper the DMCNet (Digital-Marketing-Campaign Net) is employed to score the probability of triggering a single DMC by a consumer. The score is calibrated to the probability, hence the higher the score the more probable that a consumer browsing the shop would trigger a DMC.


\subsection{Features Employed in DMCNet}
The DMCNet is a deep neural net model and its topological net structure is shown in Figure \ref{figDMCNet}. The input layer contains  four kinds of features, i.e. dense features, sparse features, target DMC threshold-discount pair and not-target DMC threshold-discount pair. The dense part  contains nine features, including order date, shop id, city id, customer id, GMV in recent 30 days, GMV in recent 60 days, GMV in recent 90 days, target DMC threshold and target DMC discount. The sparse features employed in DMCNet includes shop category, consumer age and consumer gender, which are all one-hot handled. Features specifying campaign threshold and campaign discount are divided into two parts by target and not-target. Target part colored in orange in Figure \ref{figDMCNet} contains target campaign threshold and target campaign discount, while in the not-target part many not-target campaign threshold-discount pairs colored in purple in Figure \ref{figDMCNet} are employed. Not-target campaign threshold-discount pairs are used to depict campaigns simultaneously presented to the consumers along with the target campaign threshold-discount pair. 

\subsection{DMCNet Structure }
For dense features, a three-hidden-layer fully connected forward network is employed  to learn high order features. And the result of learned vector will be concatenated with the sparse features. Here the target campaign threshold-discount pair is a scalar tuple which defines the DMC triggering threshold and the discount that the DMC triggered customer can obtain. Among the observed samples, the range of the threshold and discount varies largely, therefore in this paper the both features are monotonically encoded with a 500 length vector, which is termed as the isotonic encoding. For example, a threshold-discount pair is $<10, 1>$, meaning that in this campaign if a consumer buys $\$10$ food then the consumer will obtain a $\$1$'s discount. Then the isotonic encoding for this campaign threshold-discount pair is that $\$10$ is encoded as $(1,1,1,1,1,1,1,1,1,1,0,0,0,...,0)^{\top}_{500\times 1}$ and $\$5$ as $(1,1,1,1,1,0,0,0,...,0)^{\top}_{500\times 1}$. After isotonic encoding the target campaign threshold-discount, the 500-dimensional threshold and discount vectors are element-wise multiplied individually with the above mentioned concatenated output from dense and sparse features. The rest not-target campaign threshold-discount pairs are also handled with isotonic encoding. The encoded vectors are separately element-wise multiplied to form many output vectors with 500 dimensions.  

\subsection{DMC Embedding}
As is shown in Figure \ref{figDMCNet}, many not-target campaign threshold-discount pairs colored in purple are employed in DMCNet, whose volume varies in diverse contexts. That is for different consumers the number of all DMC threshold-discount pairs (target DMC plus not-target DMC) are different. Some consumers may be exposed with more campaigns than others. Therefore, the campaign threshold-discount pairs are specially handled as the sequential features, some sequence may be long while some may be short. Popular methods to handle the sequence data contains LSTM, GRU and Transformer-encoder. Here the LSTM with Attention is used to generate an embedding for the input not-target campaign threshold-discount pairs. This means different quantity of not-target campaign threshold-discount pairs  can be compressed into an embedding, i.e. DMC-Embedding, having the same length after LSTM-Attention module on the DMCNet. Then the DMC-Embedding will be processed by a three-hidden-layer fully connected forward net. After concatenating and again  a three-hidden-layer fully connected forward net on the Figure 1, the softmax score will be returned a score to specify the likelihood of the consumer's triggering the target DMC.

\section{4. Marketing Campaign Combination Optimization}
\subsection{Marketing Campaign Combination}
Marketing campaign response depicts  the relationship between campaign and revenue of retailer, which represented by function $f$. The input of $f$ is campaigns set up by retailer, the output of $f$ is revenue.  From business sense, submodularity of this function can be proved. First, campaigns could improve revenue by encouraging consumers to place orders, however, consumers’ willingness-to-pay is limited, which means more campaigns will make marginal effect decrease.  According to the analysis before, function $f$ obviously is submodular, which satisfies \textit{c.f.} Definition 1 and Definition 2.
 
\begin{definition}{\cite{buchbinder2018deterministicsb3}}
Given a ground set $\mathcal{N} = \{u_0, u_1,...,u_n\},\ \text{set function } f: 2^{\mathcal{N}} \xrightarrow{} R\ $ is called submodular if and only if
 $$f(A) + f(B) \ge f(A\cap B) + f(A\cup B)\ \forall A,B \subseteq \mathcal{N}$$
\end{definition}

\begin{definition}{\cite{buchbinder2018deterministicsb3}}
Given a ground set $\mathcal{N}$ = $\{u_0, u_1,...,u_n\}$, set function $f: 2^\mathcal{N} \xrightarrow{} R\ $ is called submodular if and only if
$$f(A + u) - f(A) \ge f(B + u) - f(B) \ \forall A, B \subseteq \mathcal{N}, u \in \mathcal{N} \backslash B$$
\end{definition}

More importantly, campaign combinatorial problem is not a strict monotonic increasing problem. For example, campaign\{a\}=$<39, 3>$, campaign\{b\}=$<29, 1>$, when \{b\} and  \{a\} exposed to consumers at the same time, some consumers may choose campaign \{b\} which leads the decrease of  customer price, that is $Revenue(\{a,b\}) <  Revenue(\{a \})$.

Marketing campaign combinatorial problem is described as follows,
 $\max f(A),\ s.t. \ |A| = k,\ A\subseteq \mathcal{N}$, $\mathcal{N}$ = $\{u_0, u_1,...,u_n\}.$ $k$ is the number of combination.
Therefore, campaign combinatorial problem is an unconstrained and non monotonic submodular maximization  problem.
According to the discussion, this paper proposes a solution \textit{c.f.} Figure 3 in next section .

\section{5. Recommendation Solution}
\subsection{Initializing Candidate Threshold-Discount Pairs}
\begin{figure}[htb]
\centering
\includegraphics[width=0.5\textwidth]{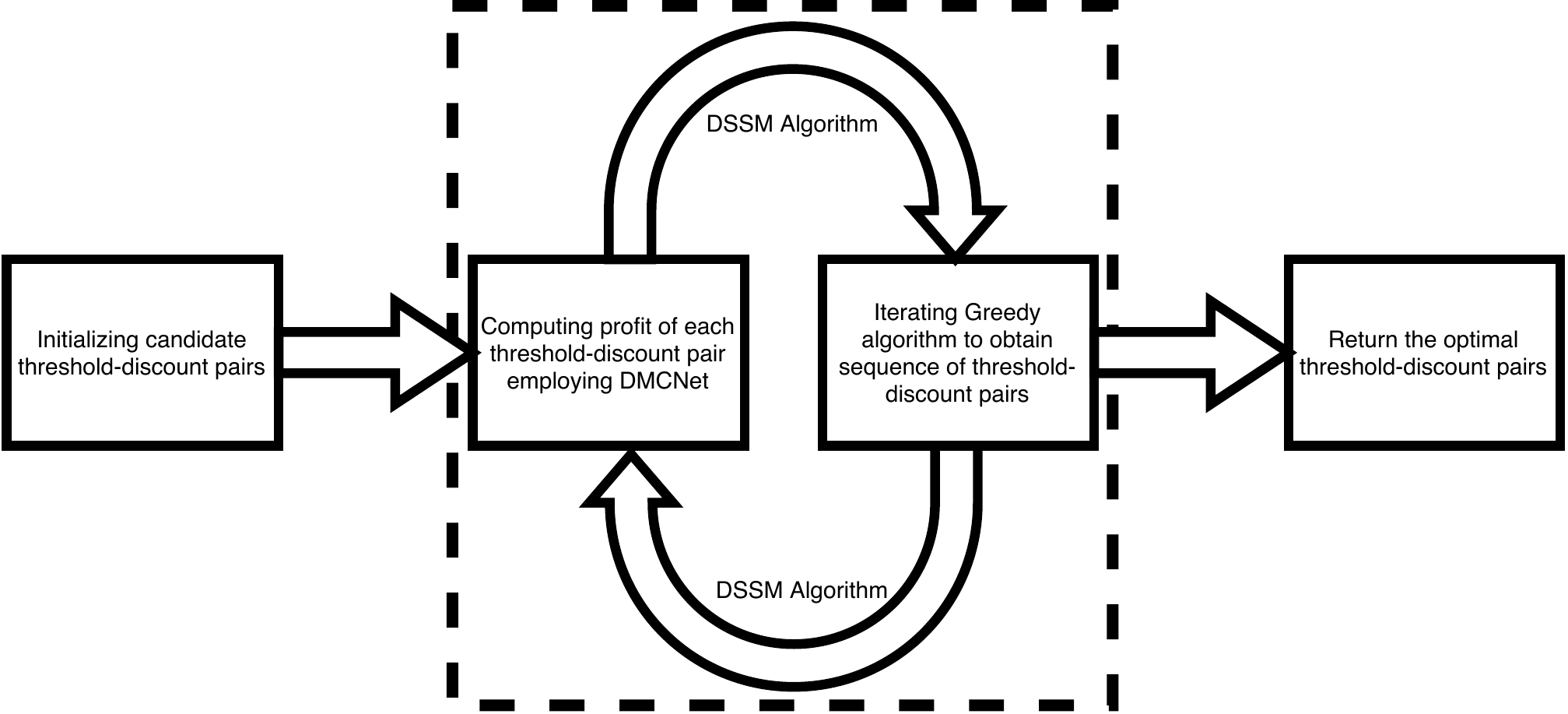}
\caption{The process graph of multiple DMCs recommending solution.}
\label{processgraph}
\end{figure}

Firstly, an iteration traverses from minimum threshold to maximum threshold, accompanying the second iteration from 0 to threshold.  As the initial candidate set of threshold-discount pairs, set $C$ contains 
all combinations of threshold and discount, \textit{c.f.} Algorithm 1.
\begin{algorithm}[h]
\caption{Initializing Basic Threshold-Discount Pairs}
\begin{algorithmic}[1]
    \STATE $C \gets \varnothing$
    \FOR{$threshold \gets threshold_{min}$ $to$ $threshold_{max}$}
        \FOR{$discount \gets 0 $ $to$ $threshold$}
            \STATE $C \gets <threshold, discount>$
        \ENDFOR
    \ENDFOR
\end{algorithmic}
\end{algorithm}

\subsection{Computing Profit of Each Threshold-Discount Pair Employing DMCNet}
Generally speaking, this paper selects consumers whose geographic location is around shops as our potential consumers.
By computing threshold and discount of each potential consumer and shop, revenue of each threshold-discount pair can be obtained. Pseudo code is given as follows \textit{c.f.} Algorithm 2. Function $F$ is the DMCNet which discussed at previous section.
\begin{algorithm}[h]
\caption{Computing Threshold-Discount-Pair Revenue }
\begin{algorithmic}[1]
    \STATE $i \gets 0;$
    \STATE $n \gets Count(C);$
    \STATE $j \gets 0;$
    \STATE $m \gets Count(user)$
    
    \FOR{$i \gets 0$ $to$ $n$}
        \STATE $Revenue_{C_i} \gets 0$
        \FOR{$j \gets 0$ $to$ $m$}
        \STATE $Revenue_{C_i} \gets Revenue_{C_i} + F(C_i, user_j, \varnothing)$
        \ENDFOR
    \ENDFOR
\end{algorithmic}
\end{algorithm}

\subsection{Iterating Greedy Algorithm to Obtain Sequence Of Threshold-Discount Pairs}
So far the set of all threshold-discount pairs and their revenue have been obtained. The set generated previously contains all threshold-discount pairs, like $<60, 1>$, $<60, 2>$, $<60, 3>$, $<50, 1>$, $<50, 2>$, $<50, 3>$ and so on, which can not be shown simultaneously. For example, in actual threshold-discount pairs, either $<60, 1>$ or $<60, 2>$ will be displayed, which means only one will be shown. Therefore, more business rules is necessary, such as, one threshold can only have one discount. The difference between adjacent thresholds must be at least $n$ , which always set $n=\$5$, finally, as threshold increases, discount must increase similarly, for example, if campaign $<50, 4>$ is contained, then the discount of 60 must be higher than $\$4 $. All of the inputs based on even common sense or business rules, will have the following advantages, firstly, it will reduce the optimal threshold-discount pair computing scale afterwards. Secondly, it makes our model more consistent with business logic. By sort set $C$ with revenue and filter set $C$ with business rules, a new set $C$ is generated. Function $fit(C_i, R, D )$ return true if $C_i$ conforms with the rules $R$ \textit{c.f.} Algorithm 3.

\begin{algorithm}[h]
\caption{Generating Fundamental Threshold-Discount-Pair Sequence}
\begin{algorithmic}[1]
    \STATE  $C \gets Sort(C, revenue)$
    \STATE  $D \gets  \varnothing$
    \FOR{$i \gets 0 $ $to$ $n$}
        \IF{$fit(C_i, R, D )$}
            \STATE $D \gets C_i + D$ 
        \ENDIF
    \ENDFOR
    \STATE $C \gets D$
\end{algorithmic}
\end{algorithm}

\subsection{Obtain Optimal Threshold-Discount Pairs By Randomized USM Algorithm}
Finally, based on the generated set of threshold-discount pairs, threshold-discount pair is added to the final set successively using greedy algorithm. For each adding action, recompute revenue to decide whether this pair should be added to final set or not.

Besides this, this paper use randomized USM algorithm (\textit{c.f.} Algorithm 4) which is proved more effectively. Not only it computes whether revenue is uplifted after one specific threshold-discount pair is added, but also computes the change of revenue after the pair is removed. 

\begin{algorithm}[h]
\caption{Randomized USM Algorithm}
\begin{algorithmic}[1]
    \STATE $X_0 \gets \varnothing, Y_0 \gets Random(C)$
    \FOR{$i = 1 $ $to$ $n$}
        \STATE $a_i \gets {F(X_{i-1} + c_i, user) - F(X_{i-1}, user)}$
        \STATE $b_i \gets {F(Y_{i-1} - c_i, user) - F(Y_{i-1}, user)}$
        \STATE ${a_i'} \gets max\{a_i,0\},  {b_i'} \gets max\{b_i,0\}$
        \STATE $with$ $probability$ $\frac{a_i'}{{a_i'} + {b_i'}} $ $do$ : $X_i \gets X_{i-1} + c_i, Y_i \gets Y_{i-1}$
        \STATE $with$ $probability$ $\frac{b_i'}{{a_i'} + b_i'} $ $do$ : $X_i \gets X_{i-1}, Y_i \gets Y_{i-1} - c_i$
    \ENDFOR
    \STATE $C \gets X_n$
\end{algorithmic}
\end{algorithm}
It can be proved that this algorithm is 1/2\ approximation algorithm for the USM problem, that means $\text{E}(f(A_k)) \geq \frac{1}{2}f(OPT)$. {\cite{buchbinder2015sb1}}




\begin{table}[htb]
\centering

\begin{tabular}{l|l|l}
\hline\hline
Search Method            & GMV                                           & Time Cost   \\
\hline\hline
Global Optimum Searching & {3,976,560} & 6.3 hours   \\
Randomized USM Searching & {2,177,239} & 20 minutes  \\
Greedy Searching         & {1,802,452} & 14 minutes  \\
\hline\hline   

\end{tabular}
\caption{GMVs and related time costs in employed searching methods. }
\label{tb6}
\end{table}

\begin{table}
\centering
\begin{tabular}{l|l|l} 
\hline
\hline
City     & Sample Size Control & Sample Size Treatment  \\ 
\hline
\hline
Jinhua   & 38                  & 17                     \\ 

Ningbo   & 42                  & 19                     \\ 

Changsha & 62                  & 20                     \\ 

Tianjin  & 158                 & 19                     \\
\hline
\hline
\end{tabular}
\caption{Sample sizes in control group and treatment group distributed in four Chinese cities.}
\label{tb1}
\end{table}

\begin{table}[htb]
\centering
\begin{tabular}{l|l|l}
\hline\hline
Model                                                                         & Train AUC & Test AUC  \\ \hline \hline
XGBoost                                                                       & 0.637     & 0.604     \\ 
DMCNet without LSTM Emb.             & 0.750     & 0.700     \\ 
DMCNet with LSTM Emb.                & 0.810     & 0.750     \\ 
DMCNet with LSTM Emb. \& Att. &0.810     &  $\mathbf{0.760}$    \\ \hline\hline
\end{tabular}
\caption{AUC comparison with four test models.}
\label{tb5}
\end{table}

\begin{table*}[htb]
\centering
\begin{tabular}{l|l|l|l|l}
\hline
\hline
Group                           & Time  & Avg Net GMV pS & Avg Order Vol. pS & Avg Net Cust. Price pS \\ \hline \hline
Treatment                       & Week1 & 1019.61            & 23.4    & 43.58     \\  
Treatment                       & Week2 & 1083.45            & 24.34   & 44.51      \\  \hline
chain ratio week1 over week2    &       & +6.26              & +4.02\%  & +2.13\%  \\ \hline
Control                         & Week1 & 1207.11            & 28.74   & 42.01      \\  
Control                         & Week2 & 1216.89            & 29      & 41.96     \\  \hline
chain ratio week1 over week2    &       & +0.81\%            & +0.9\%  & -0.12\%    \\ \hline
Ratio of Treatment over Control &       & +5.45\%            & +3.11\% & +2.25\%   \\ \hline \hline

\end{tabular}
\caption{Comparison of average net GMVs (Gross Merchandise Volume) per shop and average order volume per shop and average net customer price per shop between control group and treatment group.}
\label{tb2}
\end{table*}

\begin{table*}[htb]
\centering
\begin{tabular}{l|l|l|l}

\hline
\hline
Group                           & Time  & Net Cust. Price Aft. Triggering Camp. & Not Net Cust. Price Aft. Triggering Camp. \\ \hline \hline
Treatment                       & Week1 & 45.65     & 42.98    \\ 
Treatment                       & Week2 & 51.13     & 41.90    \\ \hline
chain ratio week1 over week2    &       & +12.02\%  & -2.52\%  \\ \hline
Control                         & Week1 & 43.64     & 41.42    \\ 
Control                         & Week2 & 41.73     & 42.05    \\ \hline
chain ratio week1 over week2    &       & -4.39\%   & +1.53\%  \\ \hline
Ratio of Treatment over Control &       & +16.41\%  & -4.05\%  \\  \hline \hline
\end{tabular}
\caption{Comparison of control group and treatment group in net customer after triggering a campaign and not net customer after triggering a campaign.  }
\label{tb3}
\end{table*}

\begin{table*}[htb]
\centering
\begin{tabular}{l|l|l|l|l}
 \hline \hline
Group                           & Time  & Ratio of Camp.-based Order Vol. Over Total Order Vol.  & App P1  & App P2   \\ \hline\hline
Treatment                       & Week1 & 22.31\%    & 5.87\%  & 24.74\%  \\ 
Treatment                       & Week2 & 28.14\%    & 6.27\%  & 26.61\%  \\ \hline
chain ratio week1 over week2    &       & 26.13\%    & 6.81\%  & 7.56\%  \\ \hline
Control                         & Week1 & 26.64\%    &7.25\%  & 27.85\%  \\ 
Control                         & Week2 & 27.94\%    & 7.18\%  & 29.44\%  \\ \hline
chain ratio week1 over week2    &       & +4.88\%    & -0.97\% & +5.71\%    \\ \hline
Ratio of Treatment over Control &       & +21.25\%   & +7.78\% & +1.85\%    \\ \hline \hline
\end{tabular}
\caption{Comparison of control group and treatment group in ratio of campaign-based order volume over total order volume, app p1 (click-through) and app p2 (conversion).}
\label{tb4}
\end{table*}

\section{6. Real Online Experiment}
\subsection{Experiment Setup}
The online experiment is an A/B-test which was performed in  diverse Chinese cities' online-shops. Treatment group and control group are selected from four Chinese cities, see Table \ref{tb1}. Totally, 65 online-shops for treatment group and 300 for control group are chosen. 

Two time windows (two weeks in total) are selected, one for control group between 2020-04-17 and 2020-04-23 and another for treatment group between 2020-04-24 and 2020-04-30.

There are many diverse types of shops in authors' online-sell-platform, including online supermarket, online convenience-store, online fruit-shop and online fresh-supermarket. In this work only  online convenience-store are tested, since this kind shops are popular in Chinese cities.

\subsection{Model Comparison}

Deep neural network tested in the experiment contains four distinct models, including XGBoost, DMCNet without embedding, DMCNet with only LSTM embedding, and DMCNet with LSTM embedding and multihead-attention.
From Table \ref{tb5}, it is outstanding that the DMCNet with LSTM embedding and multihead attention performs better than the other models. Especially, the difference between DMCNet with and without LSTM embedding is about 10 percent large. Hence it shows that the DMCNet with LSTM embedded sequence of the threshold-discount pairs is vital in improving the model performance. 

\subsection{Optimization Methods Comparison}
This paper compares three optimization methods on 100 stores data during one month. The greedy algorithm and randomized USM Algorithm and Global Search Algorithm. From Table \ref{tb6}, randomized USM algorithm obtains higher revenue than the greedy algorithm and less time than global search algorithm. 

\subsection{Evaluation Metrics}
In this experiment, eight distinct metrics have been employed, including average net GMV  per shop, average order volume per shop, and average net customer price per shop, \textit{c.f.} Table \ref{tb2}. Also net customer price after triggering a campaign and net customer price  not triggering a campaign are contained in Table \ref{tb3}. And ratio of campaign-based order volume versus total order volume, and app p1 (click-through) and app p2 (conversion) are included as evaluation metrics in Table \ref{tb4}.


\subsection{Experiment Results Discussion}
First of all, within two weeks' experiment, the real net GMV increasing rate arrived at $6.26\%$ compared to the pre-set target rate of $3\%$. 

Secondly, the daily net GMV in treatment group through week 2 is $5.45\%$ higher than it in week 1, \textit{c.f.} Table \ref{tb1}. 

Thirdly, compared with control group, the increase rate of net customer price in treatment group is higher than it in control group. 
From the perspective of net customer price after triggering a campaign and net customer price  not triggering a campaign, the former increases positively while the later negatively. This evicts that the former's effect can terminate the negative effect by the later and eventually pushes up the total net customer price, see Table \ref{tb3}. And the two weeks' chain ratios of net customer price after triggering a campaign between the control and treatment groups also demonstrates that the DMC recommendation is reasonable. 

Fourthly, in metric of campaign-based order volume over total order volume, treatment group's performance is better than the control group. This shows that the DMC recommendation indeed enlarge the order volume. Combining with the third point, it concludes that the DMC recommendation is the direct momentum to increase the net GMV in treatment group. 

Last but not the least, the click-through (APP P1) and conversion (APP P2) both increase in treatment group while not in the control group, see Table \ref{tb4}, meaning that it has positive relationship with applying multiple DMCs.

\section{7. Conclusion}
This work proposes a comprehensive solution of using digital-marketing-campaign to increase shop's GMV. First of all, the DMCNet is employed to compute the probability of triggering DMC, whose network structure is a hybrid deep learning model combining deep net and LSTM-Attention. Deep net component is used to learn the representation of user profile and contextual features and LSTM-Attention component is served as the DMC embedding generator. The combinatorial optimization is performed based on the sub-modular optimization theory to generate a set of optimal threshold-discount pairs. The real online A/B-test  on a online e-commerce platform evicts that the proposed solution increase the revenue of retailers significantly.
 
 

\end{document}